\documentclass[aps,showpacs,prb,floatfix,twocolumn,amsmath,citeautoscript]{revtex4-1}
\usepackage{graphicx}
\usepackage{amsmath}
\usepackage{amssymb}
\usepackage{bm}
\usepackage{dcolumn}
\usepackage{subfigure}
\usepackage{units}
\usepackage{hyperref}
\usepackage{longtable}
\usepackage[usenames]{color}
\hypersetup{pdfborder=0 0 0,colorlinks=true,citecolor=blue,linkcolor=blue}
\input epsf

\newcommand{\BN}{\textit{h}-BN}

\begin{document}

\title{Schottky barriers at hexagonal boron nitride$|$metal interfaces: a first principles study}
\author{Menno Bokdam}\affiliation{Faculty of Science and Technology and MESA$^{+}$ Institute for
Nanotechnology, University of Twente, P.O. Box 217, 7500 AE Enschede, The
Netherlands}
\author{Geert Brocks}\affiliation{Faculty of Science and Technology and MESA$^{+}$ Institute for
Nanotechnology, University of Twente, P.O. Box 217, 7500 AE Enschede, The
Netherlands}
\author{M. I. Katsnelson}\affiliation{Radboud University of Nijmegen, Institute for Molecules and Materials, Heijendaalseweg 135, NL-6525 AJ Nijmegen, The Netherlands}
\author{Paul J. Kelly}
\affiliation{Faculty of Science and Technology and MESA$^{+}$ Institute for
Nanotechnology, University of Twente, P.O. Box 217, 7500 AE Enschede, The
Netherlands}

\begin{abstract}
The formation of a Schottky barrier at the interface between a metal and hexagonal boron nitride ($h$-BN) is studied using density functional theory. For metals whose work functions range from 4.2 to 6.0 eV, we find Schottky barrier heights for holes between 1.2 and 2.3 eV. A central role in determining the Schottky barrier height is played by a potential step of between 0.4 and 1.8 eV that is formed at the metal$|h$-BN interface and effectively lowers the metal work function. If $h$-BN is physisorbed, as is the case on fcc Cu, Al, Au, Ag and Pt(111) substrates, the interface potential step is described well by a universal function that depends only on the distance separating $h$-BN from the metal surface. The interface potential step is largest when $h$-BN is chemisorbed, which is the case for hcp Co and Ti (0001) and for fcc Ni and Pd (111) substrates.  
\end{abstract}

\date{\today}


\maketitle

\section{Introduction}
\label{sec:intro}
Hexagonal boron nitride ($h$-BN) has a structure similar to that of graphite. 
The boron and nitrogen atoms within a single $h$-BN sheet form strong bonds on a 
hexagonal lattice, as in graphene, while the bonding between layers is weak, as 
in graphite. In contrast to graphene or graphite, $h$-BN is a large band gap 
insulator, making it a very suitable substrate and gate dielectric for 
applications in graphene electronics 
\cite{Novoselov:pnas05,Giovannetti:prb07,Dean:natn10,Xue:natm11,Decker:nanol11,
Bokdam:nanol11,Usachov:prb10,Karpan:prb11,Roth:nanol13,Bokdam:prb13}. Indeed it 
has been shown that the mobility of the charge carriers in graphene adsorbed on 
$h$-BN at low temperatures is comparable to that measured in suspended graphene 
\cite{Dean:natn10,Xue:natm11,Decker:nanol11}. At room temperature the electron 
mobility in suspended graphene is even an order of magnitude {\it lower} than in 
graphene on $h$-BN, because of scattering by thermal ripples 
\cite{Castro:prl10}; the $h$-BN substrate is atomically flat \cite{Dean:natn10} 
and ripple formation is suppressed by clamping by the van der 
Waals 
bonding between the $h$-BN substrate and the graphene layer.  

Like graphene, $h$-BN layers can be prepared by mechanical exfoliation \cite{Novoselov:pnas05}. Cleaved layers can be thinned to a single layer with a high-energy electron beam \cite{Meyer:nanol09}. Alternatively, $h$-BN layers can be grown by chemical vapor deposition (CVD) on transition metals such as Cu or Ni, using precursors such as borazine (B$_3$N$_3$H$_6$) or ammonia borane (NH$_3$-BH$_3$) \cite{Song:nanol10,Shi:nanol10}. With a proper choice of growth conditions, homogeneous ultrathin $h$-BN layers can be grown 1-5 monolayers thick. Graphene can be subsequently grown by CVD on top of $h$-BN \cite{Tang:scr13,Yang:natm13} or on top of a $h$-BN layer adsorbed on a metal substrate \cite{Usachov:prb10,Kim:nanol13,Roth:nanol13} which is ideal for field-effect devices.

Hexagonal boron nitride ($h$-BN) is an insulating material, with a measured band gap of 5.97 eV \cite{Watanabe:natm04} that is indirect \cite{Arnaud:prl06}. The direct nature of the band gap in monolayer $h$-BN\cite{Karpan08} suggests it as an interesting candidate for making ultraviolet light emitting diodes and lasers, but the size of the band gap makes it difficult to form stable $p$-$n$ junctions or ohmic metal contacts. It has been suggested that $h$-BN is a material with a negative electron affinity \cite{Loh:apl99}, which would make obtaining a low barrier contact for electrons extremely challenging. At the same time this would improve the chances for making low barrier contacts for holes. Having low barrier metal contacts is advantageous for using $h$-BN as a semiconductor but disadvantageous for graphene devices where $h$-BN is used as an insulator. 

In this paper we use density functional theory (DFT) calculations to determine the hole Schottky barrier height $\Phi_p$ between a metal contact and $h$-BN for a series of metals whose work functions range from 4.2 to 6.0 eV. Substantial Schottky barrier heights ($1.2 < \Phi_p < 2.3$ eV) are found in all cases. The relationship between the Schottky barrier heights and the clean metal work functions is not simple. A key role is played by a potential step at the metal$|h$-BN interface, which originates from an interface dipole layer that is formed when $h$-BN is adsorbed on a metal surface. This potential step effectively lowers the metal work function and increases the Schottky barrier height for holes. For weak adsorption (physisorption), we find that the size of the interface dipole can be described by a universal function that only depends on the metal$|h$-BN bonding distance, and not on the metal.  

Special attention is devoted to the case where the adsorption energies are weak and there is a lattice mismatch between $h$-BN and the metal surface. Under these circumstances, we expect that a superstructure will be formed with a periodically modulated interface dipole and potential step that have the same periodicity as the superstructure.

\section{Method and Computational details}
\label{sec:method}

We use density functional theory (DFT) to calculate ground state energies and optimized geometries with a projector augmented wave (PAW) basis set \cite{Blochl:prb94b,Kresse:prb99} as implemented in \textsc{vasp} \cite{Kresse:prb93,Kresse:prb96}. The interfaces are modeled using $h$-BN adsorbed on finite slabs of metal, six close-packed atomic layers thick. This bilayer structure is repeated periodically and separated from its images by a thick vacuum region. A dipole correction is applied to avoid spurious interactions between the  periodic images \cite{Neugebauer:prb92}. The plane wave kinetic energy cutoff is set at 400 eV. We apply a $24\times 24$ \textbf{k}-point grid to sample $1\times 1$ and $\sqrt{3}\times \sqrt{3}$ surface Brillouin zones (BZ) and use the tetrahedron scheme for BZ integrations \cite{Blochl:prb94a}. The electronic self-consistency criterion is set to $10^{-7}$ eV. Interface geometries are relaxed until the total energy is converged to within $10^{-6}$ eV. The boron and nitrogen atoms in the {\it h}-BN slab, and the top two atomic layers of the metal surface are allowed to relax during the geometry optimization. The other four metal layers are kept fixed in their crystal structure. For the $5\times 5$ Ti surface supercell we use a $3\times 3$ \textbf{k}-point grid during geometry relaxation and a $9\times 9$ grid to converge the charge density of the optimized geometry.

The interaction in weakly (van der Waals) bonded systems is difficult to 
describe within DFT. Commonly used generalized gradient approximation (GGA) 
functionals, such as PW91 or PBE, incorrectly predict essentially no bonding 
between $h$-BN or graphene layers \cite{Giovannetti:prb07,Sachs:prb11}, as well 
as no bonding between $h$-BN or graphene and transition metal (111) surfaces 
\cite{Laskowski:prb08,Khomyakov:prb09,Olsen:prl11}. The local density 
approximation (LDA) is empirically found to give a reasonable description of the 
bonding (geometry and energy) between $h$-BN 
layers,\cite{Giovannetti:prl08,Khomyakov:prb09,Bokdam:prb13}
between $h$-BN or graphene and metal (111) surfaces and between a $h$-BN and a 
graphene layer \cite{Giovannetti:prb07,Bokdam:nanol11,Bokdam:prb13}. 
Moreover, very accurate AC-FDT-RPA (adiabatic-connection 
fluctuation-dissipation theory in the random phase approximation) binding energy 
calculations for graphene on $h$-BN by Sachs \emph{et al.}\cite{Sachs:prb11} are 
in good agreement with the LDA.\cite{Bokdam:prb14a}  
In general, however, the LDA is known to overestimate 
chemical bonding, and it does not capture van der Waals interactions properly. 

In this paper we use one of the recently developed and implemented van der Waals density functionals (vdW-DF) \cite{Dion:prl04,Thonhauser:prb07,Klimes:prb11}, and compare the results to those obtained with LDA. In the vdW-DF, the exchange-corrrelation functional is split up as $E_{\rm xc}=E_{\rm x}+E_{\rm c}^{\rm vdW}+E_{\rm c}^{\rm loc}$, where $E_{\rm c}^{\rm vdW}$ describes non-local electron-electron correlations and $E_{\rm c}^{\rm loc}$ local correlations. For $E_{\rm c}^{\rm vdW}$ and $E_{\rm c}^{\rm loc}$ we use the vdW kernel developed by Dion \emph{et al.} \cite{Dion:prl04} and the LDA correlation \cite{Ceperley:prl80}, respectively. For the exchange part $E_{\rm x}$, we use the optB88 functional \cite{Klimes:prb11}. The resulting optB88-vdW-DF functional gives a satisfactory description of the lattice parameters and binding energy of graphite, as well as of the structures and energetics of Li intercalation in graphite \cite{Hazrati13}.

For the systems we will be studying, it is not clear whether the LDA or the vdW-DF gives a better description of reality. Most of our results were obtained with the LDA but the vdW-DF corrections to the most important results are also given. Unless stated otherwise, results were obtained with the LDA.

\section{Results}  
\label{sec:results}

\subsection{Metal$|h$-BN structures and bonding} 
\label{ssec:mbinding}

\begin{figure}[b]
\includegraphics[width=8.5cm]{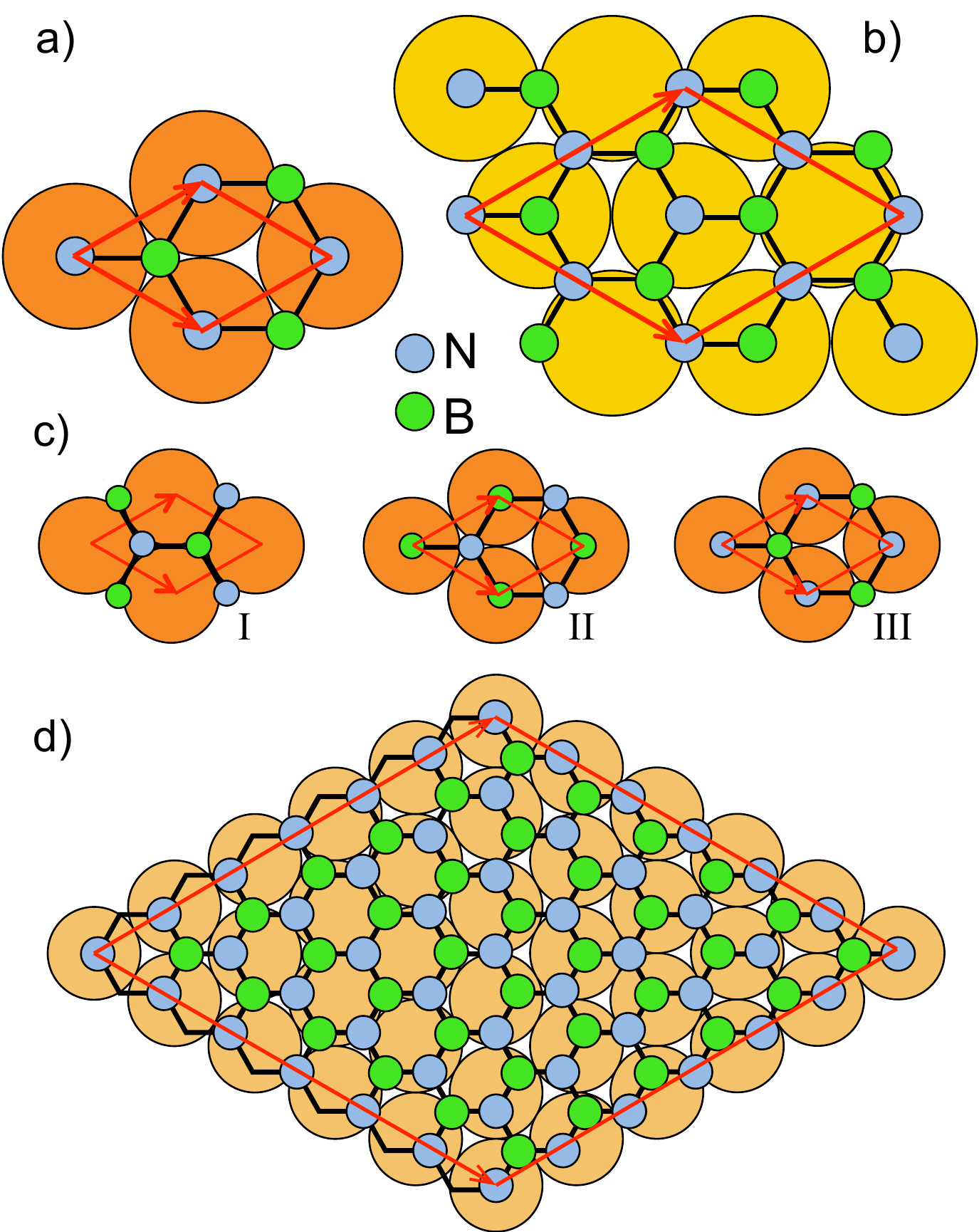}
\caption{(Color online) The binding configurations of $h$-BN on a) Cu(111), Ni(111), Co(0001) and b) Pt, Pd, Au, Ag, and Al (111). Panel a) shows a 1$\times$1 $h$-BN unit cell on top of a $1 \times 1$ metal surface cell while panel b) shows a 2$\times$2 $h$-BN unit cell on top of a $\sqrt{3} \times \sqrt{3}$ metal surface cell. c) Three high symmetry configurations of 1$\times$1 $h$-BN on a $1 \times 1$ metal surface  referred to in Table \ref{tbCuBN}; a) is the same as configuration III. d) Binding configuration of a $6 \times 6$ unit cell of $h$-BN on top of a $5 \times 5$ Ti(0001) surface cell. 
} 
\label{figB}
\end{figure}

\begin{table*}[t]
\caption{Calculated (LDA) and experimental potential steps $\Delta_{\rm M|BN}$ at the metal$|h$-BN interface; (average) equiilibrium distance $d_\mathrm{eq}$ between the metal surface and the $h$-BN plane, using the structures shown in Fig.~\ref{figB}, starting from the optimized in-plane $h$-BN lattice constant of 2.49 \AA, and scaling the in-plane metal lattices accordingly, see also the Appendix; binding energies $E_{\rm b}$ per BN; calculated and experimental work functions $W_{\rm M|BN}$ of metal covered with a single sheet of $h$-BN and $W_{\rm M}$ of the clean metal surface; the position $\Phi^*_p$ of the Fermi level with respect to the top of the $h$-BN valence band. The last two columns show the Schottky barrier height $\Phi_p$ for holes calculated for several layers of $h$-BN, see Eq. (\ref{eq:schottky}) with $E_v=6.09$ eV and the Schottky barrier height $\Phi_n$ for electrons, see Eq. (\ref{eq:nschot}), where we have used the experimental band gap $E_g = 5.97$ eV.\cite{Watanabe:natm04} 
The interface dipole can extend further than the first $h$-BN layer requiring an additional correction $\Delta_{\rm BN}$, that is found to be $\sim 0.0$~eV for physisorption and $\sim 0.1$~eV for chemisorption, see Sect.~\ref{ssec:levels}.
}
\begin{ruledtabular}
\begin{tabular}{lccccccclc|cc}
 M  & $\Delta_{\rm M|BN}$
           & $\Delta_{\rm M|BN}^{\rm exp} $   
                         & $d_{\rm eq} $
                                & $E_{\rm b}$ 
                                           &  $W_{\rm M|BN} $ 
                                                  &$W_{\rm M|BN}^{\rm exp} $  
                                                                  &  $W_{\rm M} $ 
                                                                         &$W_{\rm M}^{\rm exp} $ 
                                                                                          &$\Phi^*_{p}$
                                                                                                 & $\Phi_{p}$ 
                                                                                                          & $\Phi_{n}$\\
 \hline
 Co & 1.80 &             & 1.92 & $-0.583$ & 3.72 &               & 5.52 & 5.55$^{\rm d}$ & 4.56 & $2.27$ & 3.70 \\
 Ni & 1.73 & 1.5-1.8$^a$ & 1.96 & $-0.430$ & 3.79 & 3.6$^{\rm h}$ & 5.52 & 5.35$^{\rm e}$ & 4.39 & $2.20$ & 3.77 \\
 Ti & 0.78 &             & 2.17 & $-0.305$ & 3.65 &               & 4.43 & 4.58$^{\rm i}$ & 3.63 & $2.34$ & 3.63 \\
 Pd & 1.25 & 1.3$^b$     & 2.47 & $-0.163$ & 4.28 & 4.0$^{\rm h}$ & 5.53 & 5.6 $^{\rm e}$ & 2.97 & $1.71$ & 4.26 \\
 Cu & 1.18 & 0.8-1.1$^c$, 0.24$^g$ 
                         & 2.97 & $-0.087$ & 3.99 &               & 5.17 & 4.98$^{\rm e}$ & 2.47 & $2.10$ & 3.87 \\
 Pt & 1.04 & 0.9$^b$     & 3.04 & $-0.100$ & 4.94 & 4.9$^{\rm h}$ & 5.98 & 6.1 $^{\rm f}$ & 1.25 & $1.15$ & 4.82 \\
 Ag & 0.83 &             & 3.20 & $-0.070$ & 4.00 &               & 4.83 & 4.74$^{\rm e}$ & 2.18 & $2.09$ & 3.88 \\
 Au & 0.79 &             & 3.24 & $-0.075$ & 4.76 &               & 5.55 & 5.31$^{\rm e}$ & 1.40 & $1.33$ & 4.64 \\
 Al & 0.41 &             & 3.55 & $-0.052$ & 3.84 &               & 4.25 & 4.24$^{\rm e}$ & 2.34 & $2.25$ & 3.72 \\
    & (eV) &  (eV)      &(\AA{})&  (eV)    & (eV) &  (eV)         & (eV) & (eV)           & (eV) &  (eV)  & (eV) \\
\end{tabular}
\end{ruledtabular} 
\newline
$\rm ^a$ Refs. \onlinecite{Preobrajenski:ss05,Nagashima:ss96,Grad:prb03,Leuenberger:prb11}, 
$\rm ^b$ Ref. \onlinecite{Nagashima:ss96}, 
$\rm ^c$ Ref. \onlinecite{Joshi:nanol12},
$\rm ^d$ Ref. \onlinecite{Vaara:ss98},
$\rm ^e$ Ref. \onlinecite{Michaelson:jap77}, 
$\rm ^f$ Ref. \onlinecite{Derry:prb89},
$\rm ^g$ Ref. \onlinecite{Preobrajenski:ss05},
$\rm ^h$ Ref. \onlinecite{Nagashima:prl95},
$\rm ^i$ Ref. \onlinecite{Jonker:prb81}.
\label{tbeq}
\end{table*}

\begin{table*}
\caption{As Table \ref{tbeq} but with all calculated values obtained using optB88-vdW-DF so that $a_{\rm hex} = 2.510$ \AA. The Schottky barrier height for holes, $\Phi_{p}$, see Eq. (\ref{eq:schottky}), is calculated with $E_v$=6.04 eV.  \cite{fn2}
}
\begin{ruledtabular}
\begin{tabular}{lccccccclc|cc}

 M  & $\Delta_{\rm M|BN}$  &  $\Delta_{\rm M|BN}^{\rm exp} $   & $d_{\rm eq} $& $E_{\rm b}$ &  $W_{\rm M|BN} $ &$W_{\rm M|BN}^{\rm exp} $  &  $W_{\rm M} $ &$W_{\rm M}^{\rm exp} $ &$\Phi^*_{p}$& $\Phi_{p}$ & $\Phi_{n}$\\
 \hline
 Co & 1.87 &             & 2.02 & $-0.348$ & 3.55 &               & 5.42 & 5.55$^{\rm d}$ & 4.38 & $2.39$ & 3.58\\
 Ni & 1.75 & 1.5-1.8$^a$ & 2.12 & $-0.195$ & 3.65 & 3.6$^{\rm h}$ & 5.40 & 5.35$^{\rm e}$ & 4.03 & $2.29$ & 3.68\\
 Pd & 0.87 & 1.3$^b$     & 3.01 & $-0.170$ & 4.55 & 4.0$^{\rm h}$ & 5.48 & 5.6$^{\rm e}$  & 1.94 & $1.33$ & 4.64\\
 Cu & 0.82 & 0.8-1.1$^{c}$, 0.24$^{g}$ 
                         & 3.27 & $-0.140$ & 4.28 &               & 5.10 & 4.98$^{\rm e}$ & 1.61 & $1.76$ & 4.21\\
 Pt & 0.83 & 0.9$^b$     & 3.27 & $-0.160$ & 5.13 & 4.9$^{\rm h}$ & 5.96 & 6.1$^{\rm f}$  & 0.89 & $0.91$ & 5.06\\
 Ag & 0.71 &             & 3.37 & $-0.131$ & 4.11 &               & 4.82 & 4.74$^{\rm e}$ & 1.89 & $1.93$ & 4.04\\
 Au & 0.66 &             & 3.37 & $-0.144$ & 4.92 &               & 5.58 & 5.31$^{\rm e}$ & 1.19 & $1.12$ & 4.85\\
 Al & 0.38 &             & 3.67 & $-0.116$ & 3.82 &               & 4.20 & 4.24$^{\rm e}$ & 2.21 & $2.22$ & 3.75\\
    & (eV) &  (eV)      & (\AA{})&  (eV)   & (eV) &  (eV)         & (eV) & (eV)           & (eV) & (eV)   & (eV)\\
\end{tabular}
\end{ruledtabular} 
\label{tbeqvdW} 
\end{table*}

We first generate equilibrium structures for a monolayer of $h$-BN on the different transition metal surfaces. As indicated in Fig.~\ref{figB}, a $1 \times 1$ $h$-BN cell fits rather well on a $1 \times 1$ surface unit cell of Co(0001), Ni(111), and Cu(111), whereas a $2 \times 2$ $h$-BN cell can be used to match a $\sqrt{3} \times \sqrt{3}$ surface cell of Al, Pd, Ag, Pt, and Au(111). Using these cells, the mismatch between the in-plane metal and $h$-BN lattices is then $\lesssim 1$\% in most cases. The largest mismatch is 3.5\% and 4.6\% for Pt and Pd, respectively. The mismatch between the primitive $h$-BN and the Ti(0001) lattices is large, but a $6 \times 6$ supercell of $h$-BN placed on top of a $5 \times 5$ surface unit cell of Ti(0001) results in a mismatch of only 1.3\%.    

To cope with the residual mismatch, we scale the in-plane lattice constant of 
the metal to match the optimized $h$-BN lattice constants of $a_{\rm 
hex}=2.490$~{\AA} (LDA) or $2.510$~{\AA} (vdW-DF), both of which are close to 
the experimental lattice constant of $a_{\rm hex}^{\rm exp}=2.504$~{\AA} 
\cite{Lynch:jcp66}. In a previous DFT calculation for $h$-BN adsorbed on 
transition metals, $1 \times 1$ cells were used with $h$-BN scaled to the metal 
lattice constants \cite{Laskowski:prb08}. This led to overstrained $h$-BN, 
which, depending on the size of the lattice mismatch, can result in very 
unrealistic bonding \cite{WangQJ:prl09}. A small stretching of the 
in-plane metal lattice constant has a moderate effect, see the Appendix.
  
The energetically most favorable binding configurations are shown in Fig.~\ref{figB}. In the $1 \times 1$ metal surface cells, the most favorable position for the nitrogen and boron atoms are top and hollow sites, respectively, see Fig.~\ref{figB}(a) \cite{Grad:prb03,Huda:prb06,Laskowski:prb08,Bokdam:prb13}. In the $\sqrt{3} \times \sqrt{3}$ metal surface cells, three boron and three nitrogen atoms are adsorbed on bridge sites, and one each on top sites, see Fig.~\ref{figB}(b).

The results of the LDA calculations are shown in Table~\ref{tbeq}. The equilibrium separations $d_{\rm eq}$ are suggestive of two different bonding situations. $h$-BN is chemisorbed on hcp Co and Ti (0001), and on fcc Ni and Pd (111) surfaces with $d_{\rm eq}<2.5$ \AA,  whereas it is physisorbed on Cu, Pt, Ag, Au and Al (111) surfaces with $d_{\rm eq}>3.0$ \AA. This difference is also apparent in the bonding energies $E_{\rm b}$ (defined as the difference between the total energies of the combined system minus those of the separate systems). The physisorbed $h$-BN layer has a binding energy ranging between $-50$ and $-100$ meV per BN, while the binding energy of chemisorbed $h$-BN ranges from $-150$ to $-600$ meV. The $h$-BN monolayer buckles slightly when chemisorbed, with a height difference between boron and nitrogen on Co(0001) of 0.13 \AA{}. The corrugation of the $h$-BN sheet is largest on Ti; in the Ti supercell a `wavy' $h$-BN layer is formed in which the difference between the highest and lowest nitrogen atoms is 0.7 \AA{}. The buckling is insignificant when $h$-BN is physisorbed.

The adsorption of $h$-BN on metal substrates is quite similar to that of graphene, that is also found to be chemisorbed on Co, Ti, Ni and Pd and physisorbed on Cu, Pt, Ag, Au and Al \cite{Khomyakov:prb09}. The bonding distances $d_{\rm eq}$ of $h$-BN and graphene on these metal surfaces are within 10\% of one another, and their sequence is the same, i.e., $d_{\rm eq}$ is smallest for Co and largest for Al. So in spite of the large difference in electronic structures of graphene and $h$-BN (conductor vs insulator), their bonding to metals is very similar.

Table \ref{tbeqvdW} shows the results obtained with the optB88-vdW-DF functional. Comparing the bonding distances to the corresponding LDA results in Table \ref{tbeq} we see that the classification into chemisorbed and physisorbed $h$-BN layers is the same as in the LDA. The equilibrium bonding distances obtained with optB88-vdW-DF are, however, somewhat larger. For most metals $d_{\rm eq} $ is larger by a moderate 0.1-0.2 \AA, but for Cu it is $0.3$ \AA\ larger compared to LDA, and for Pd, $0.5$ \AA\ larger. The bonding to Pd is somewhat special, as demonstrated also by the binding energies. 

The absolute binding energy of $h$-BN on Co and Ni obtained with optB88-vdW-DF 
is 0.24 eV 
per BN lower than that obtained with LDA. This is consistent with the 
overbinding one expects LDA to give for chemisorption. In contrast, the 
absolute binding 
energy of $h$-BN on Cu, Pt, Ag, Au, and Al obtained with optB88-vdW-DF is 
0.06-0.07 eV/BN higher than that obtained with LDA. This is consistent with an 
improved description of van der Waals interactions in the vdW-DF, which is 
important for physisorption. The binding energies of $h$-BN on Pd obtained with 
optB88-vdW-DF and LDA are practically the same, which is consistent with the 
bonding being on the borderline between chemisorption and physisorption. Judging 
from the bonding distances, LDA puts Pd on the chemisorption side of that 
border, and  optB88-vdW-DF puts it more on the physisorption 
side.\footnote{In the chemical bonding regime the (semi)local part 
of the vdW functional consisting of GGA exchange and LDA correlation parts plays 
a dominant role. Whereas a pure LDA functional tends to overbind, a vdW 
functional might well underbind in this regime, depending on the system being 
studied.}

\begin{figure} [t]
\includegraphics[width=8.5cm]{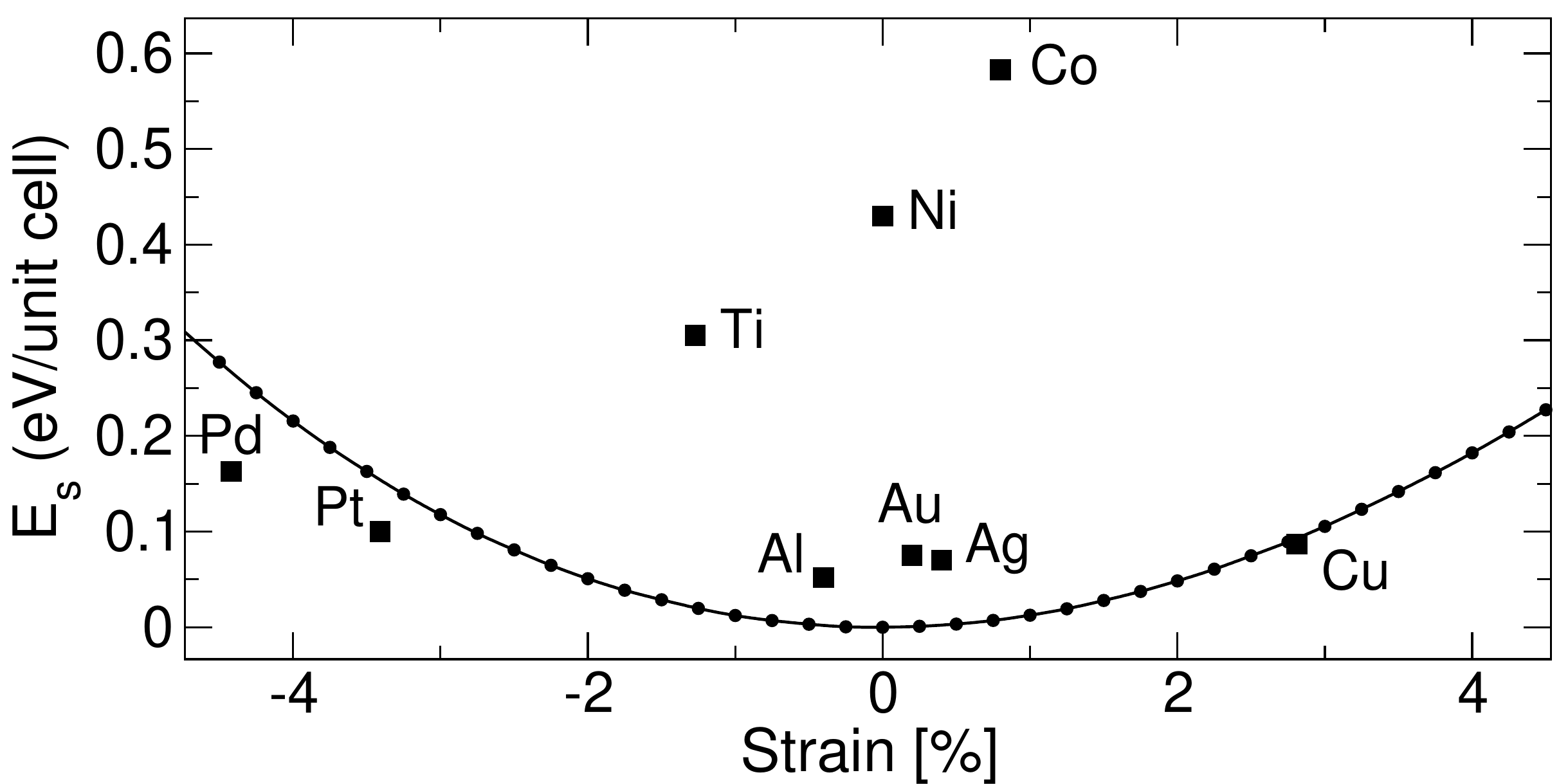}
\caption{Strain energy $E_s$ in eV versus $h$-BN strain in \%. The squares show the LDA binding energies $|E_\mathrm{b}|$ of $h$-BN on different metals from Table~\ref{tbeq}, versus the lattice mismatch between $h$-BN and the metal substrate in the supercells of Figure \ref{figB}.} 
\label{figSTRAIN}
\end{figure}

First-principles calculations are practicable only for commensurable structures with reasonably sized unit cells. Whether or not an adsorbate is commensurable with a substrate depends on the outcome of a competition between adsorbate strain energy and binding energy to the substrate; the interaction between the $h$-BN adsorbate and the metal substrate in the present case may be too weak to make commensurability energetically favourable when the lattice mismatch is too large. To address this question, we plot the strain energy of $h$-BN versus strain in Fig.~\ref{figSTRAIN}. The LDA binding energies of Table \ref{tbeq} are plotted versus the lattice mismatch in the same figure (the vdW-DF binding energies lead to the same qualitative conclusions). We expect that a metal$|h$-BN structure will be commensurable if the energy gained by binding outweighs the energy cost of straining.

For $h$-BN on Co(0001) and Ni(111), the lattice mismatch is small and the binding energies are much higher than the strain energy. Therefore it is very likely that $h$-BN will form a commensurable structure on Co(0001) and Ni(111) with a unit cell as in Fig.~\ref{figB}(a). The uniqueness of this structure would then promote the growth of large areas of defect-free $h$-BN. 

\begin{table}[b]
\caption{LDA binding energies, equilibrium distances and potential steps  for the three configurations of $h$-BN on Cu(111) illustrated in Fig.~\ref{figB}(c).}
\begin{ruledtabular}
\begin{tabular}{cccc}
Configuration & $E_{\rm b}$ (eV) & $d_{\rm eq}$ (\AA{}) & $\Delta_{\rm M|BN}$ (eV) \\
 \hline\hline
 I & -0.062 & 3.27 &  0.72 \\
 II & -0.064 & 3.25 &  0.73 \\ 
 III & -0.087 & 2.97 &  1.18 \\
\end{tabular} 
\end{ruledtabular}
\label{tbCuBN}
\end{table}

If $h$-BN is grown on the (111) surfaces of Al, Au and Ag or on the Ti(0001) surface the situation is more complicated. On the one hand, the lattice mismatch is small and the binding energy is an order of magnitude larger than the strain energy. Commensurable structures with unit cells as in Figs.~\ref{figB}(b,d) should therefore be stable. On the other hand, these unit cells can accommodate several structures that have the same energy. For instance, shifting the $h$-BN overlayer in Fig.~\ref{figB}(b) by $(\frac{1}{3},\frac{1}{3})$ or rotating it by 60$^\mathrm{o}$ with respect to the center of the cell will result in such structures. This means that during growth different domains of $h$-BN can form. An experimental LEED study of $h$-BN growth on Ag(111) indeed shows that $h$-BN domains are formed \cite{Muller:prb10}. Quite likely the same will happen for $h$-BN grown on Au and Al(111).

The lattice mismatch for $h$-BN on Cu, Pd and Pt(111) is large and is likely to 
lead to incommensurable structures. As Figure \ref{figSTRAIN} shows, the strain 
energy needed to compress (on Pd, Pt) or stretch (on Cu) the $h$-BN overlayer is 
comparable to the binding energy. Moreover, some of the binding energy is gained 
even if the atoms of the $h$-BN overlayer are not in their most favorable 
adsorption positions. The LDA binding  energies of the three configurations of 
$h$-BN on Cu(111) illustrated in Fig.~\ref{figB}(c), are given in Table 
\ref{tbCuBN}. These suggest that it is more favorable to adsorb $h$-BN in its 
equilibrium (unstrained) structure, rather than match it 1:1 with the Cu surface 
unit cell. The result is an incommensurable structure. 

The difference in bonding distance between the three configurations in Fig.~\ref{figB}(c) is 0.30 \AA{}. We expect this will be reflected in height variations within a $h$-BN layer on Cu(111). An incommensurable structure generally leads to the observation of moir\'e patterns in the adsorbed layer. Indeed recent STM experiments have shown such patterns in $h$-BN on Cu(111) \cite{Joshi:nanol12}. The effect of incommensurability on the electronic properties will be discussed further in Sec.~\ref{ssec:incom}.  

\subsection{Metal$|h$-BN interface dipole}

\begin{figure*}
\includegraphics[width=17.5cm]{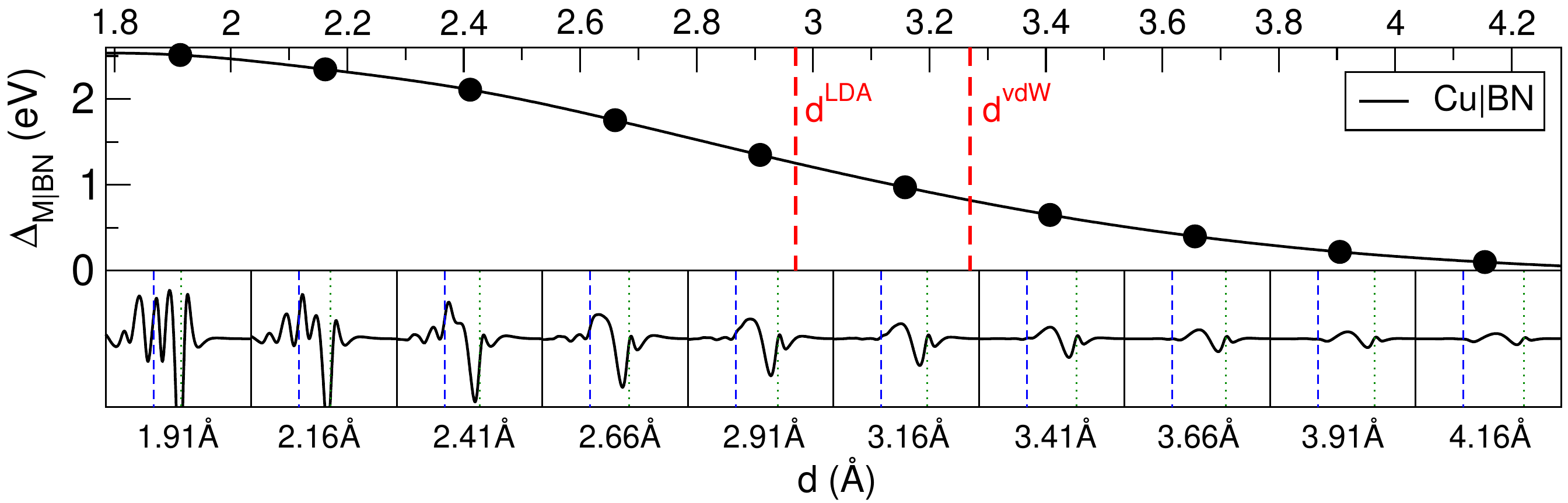}
\caption{(Color online) Top: size of the potential step $\Delta_{\rm M|BN}$ at a Cu(111)$|h$-BN interface as a function of the separation $d$ of the $h$-BN sheet from the Cu surface. The vertical (red) dashed lines indicate the calculated equilibrium separations, $d_{\rm eq}^{\rm LDA}= 2.97$~\AA\ and $d_{\rm eq}^{\rm vdW-DF}= 3.27$~\AA. 
Bottom: plane-averaged electron density difference $\Delta \overline{n}(z)$ at selected distances $d$. The positions of the top metal layer and of the $h$-BN layer are indicated by the (blue) dashed lines and the (green) dotted lines, respectively.}
\label{figD}
\end{figure*}

It has been experimentally shown that the work function of a metal surface can be substantially modified by deposition of a $h$-BN monolayer. The changes reported range from 0.9 eV for $h$-BN on Pt(111) \cite{Nagashima:ss96} to 1.8 eV for $h$-BN on Ni(111) \cite{Nagashima:ss96,Grad:prb03}. In our own work on metal$|h$-BN$|$graphene (M$|$BN$|$Gr) stacks, we showed that a dipole layer is formed at the interface between the metal and $h$-BN \cite{Bokdam:nanol11,Bokdam:prb13}. The dipole layer gives rise to a step $\Delta_{\rm M|BN}$ in the electrostatic potential at the interface, which effectively lowers the work function of the system. This potential step is of key importance for the Schottky barrier height that we will discuss in the next section.

The potential step corresponding to the dipole layer can be determined by calculating the difference between the (planar averaged) electrostatic potential sufficiently far away from the metal$|h$-BN slab and that of the clean relaxed metal surface, assuming a common Fermi level \cite{Rusu:prb10}. This is equivalent to defining the interface potential step as the difference between the work functions of the clean metal surface and the $h$-BN covered surface, 
\begin{equation}
\Delta_{\rm M|BN}=W_{\rm M}-W_{\rm M|BN}.
\label{eq:mbn}
\end{equation}
Alternatively, the interface potential step can be calculated from the electron density difference, defined by subtracting the densities of the isolated metal and $h$-BN slabs from that of the combined metal$|h$-BN slab
\begin{equation}
\label{deltan}
\Delta n(\mathbf{r})=n_{\rm M|BN}(\mathbf{r})-n_{\rm M}(\mathbf{r})-n_{\rm BN}(\mathbf{r}).
\end{equation}
As the system as a whole is neutral and $\Delta n(\mathbf{r}) \rightarrow 0$ for $\mathbf{r}$ sufficiently far from the metal--$h$-BN interface, solving the Poisson equation with $\Delta n(\mathbf{r})$ as source gives the potential step in terms of the interface dipole\cite{Rusu:prb10}
\begin{equation}
\Delta_{\rm M|BN}=\frac{e^2}{\epsilon_{0}}\int{\Delta \overline{n}(z)z \, dz}, \label{eq:mbn2}
\end{equation}
where $z$ is the direction normal to the metal$|h$-BN interface, and 
\begin{equation}
\label{deltanz}
\Delta \overline{n}(z)=\frac{1}{A}\iint \Delta n(\mathbf{r})\, dx dy,
\end{equation}
is the electron density difference averaged over a plane and $A$ is the surface area of the supercell. Numerically, the results of Eqs.~(\ref{eq:mbn}) and (\ref{eq:mbn2}) are within a few meVs of one another.

The potential steps $\Delta_{\rm M|BN}$ calculated with the LDA and the vdW-DF functionals are listed in Tables \ref{tbeq} and \ref{tbeqvdW} for all the metal$|h$-BN structures discussed in this paper. The numbers in Table \ref{tbeq} are within 0.2 eV of those in Table \ref{tbeqvdW}, except for Cu and Pd, where the differences are 0.36 eV and 0.38 eV, respectively. These differences cannot be ascribed directly to differences between the functionals. In fact, as a function of the metal--$h$-BN separation $d$, both functionals give the same values of $\Delta_{\rm M|BN}(d)$ within 0.05 eV \cite{Bokdam13}. However, they predict slightly different equilibrium bonding distances $d_\mathrm{eq}$, see Tables \ref{tbeq} and \ref{tbeqvdW}, and thus different values of $\Delta_{\rm M|BN}(d_\mathrm{eq})$. The difference is largest for the two metals for which the difference between the $d_\mathrm{eq}$ predicted by the two functionals is largest, i.e., for Cu and Pd.

Experimental results for $\Delta_{\rm M|BN}$, where available, are also given in Tables \ref{tbeq} and \ref{tbeqvdW}. The agreement with the calculated values is generally quite good, both for the LDA and the vdW-DF results. For $h$-BN on Cu(111) two quite different results have been reported for the potential step, i.e., 0.24 eV in Ref.~\onlinecite{Preobrajenski:ss05} and 0.8-1.1 eV in Ref.~\onlinecite{Joshi:nanol12}. The calculated results suggest that the latter value is more likely to represent well-ordered $h$-BN on clean Cu(111). A possible origin of the 0.3 eV spread in the measured results of Ref.~\onlinecite{Joshi:nanol12} is discussed in Sec.~\ref{ssec:incom}. 

The potential step of 1.25 eV for Pd calculated with LDA is much closer to the experimental value of 1.3 eV \cite{Nagashima:ss96} than the vdW-DF value of 0.87 eV. Again this does not directly reflect the difference between the two functionals, but rather the difference between the equilibrium bonding distances $d_\mathrm{eq}$ these functionals predict. As discussed in the previous section, LDA gives a more chemisorbed $h$-BN layer with a shorter bonding distance ($d_\mathrm{eq} \approx 2.5$ \AA) than vdW-DF ($d_\mathrm{eq} \approx 3.0$ \AA). Comparison of the calculated and experimental values of the potential steps suggests that the shorter bonding distance is more likely.

A typical distance dependence of the potential step $\Delta_{\rm M|BN}(d)$ is shown in Fig. \ref{figD} for the Cu$|h$-BN interface. Obviously $\Delta_{\rm M|BN}$ is very sensitive to $d$. The functional dependence can be understood in terms of exchange repulsion between the metal surface and the $h$-BN layer at intermediate distances, which is strongly modified by chemical interactions at shorter distances \cite{Bokdam13}. The separation into two regimes is illustrated in Fig.~\ref{figD} by plotting $\Delta \overline{n}(z)$ for different values of the distance between the Cu(111) surface and $h$-BN sheet. For distances $d \gtrsim 3.0$ {\AA}, $\Delta \overline{n}(z)$ displays the pattern of a simple dipole, with accumulation of electrons near the Cu(111) surface and a concommittant depletion near the $h$-BN plane. Such an accumulation/depletion pattern is also called the pillow effect or the pushback effect because it appears as if adsorption of the overlayer pushes electrons into the metal substrate. It can be shown that exchange (Pauli) repulsion provides the dominant contribution to the dipole in this distance regime \cite{Bokdam13}.

At smaller separations, $d<3.0$~\AA, the pattern of $\Delta \overline{n}(z)$ becomes more complicated than that of a simple dipole. $\Delta \overline{n}(z)$ shows oscillations in the metal and this strong perturbation of the electron density in the metal is indicative of the formation of chemical bonds (or anti-bonds). Qualitatively, it exhibits a similar dependence on distance $d$ for all metal$|h$-BN interfaces. For metals where the equilibrium separation $d_\mathrm{eq}<3.0$ {\AA}, one observes a pattern with strong oscillations in the metal \cite{Bokdam:prb13}. These are the metals Co, Ti, Ni, and Pd on which $h$-BN is chemisorbed according to Table~\ref{tbeq}. In contrast, for metals where the equilibrium separation $d_\mathrm{eq}\gtrsim3.0$ {\AA}, one observes the pattern of a simple pushback dipole \cite{Bokdam:prb13}. These are the metals Al, Cu, Ag, Au, and Pt on which $h$-BN is physisorbed according to Table \ref{tbeq}.

\begin{figure}[!b]
\includegraphics[width=8.5cm]{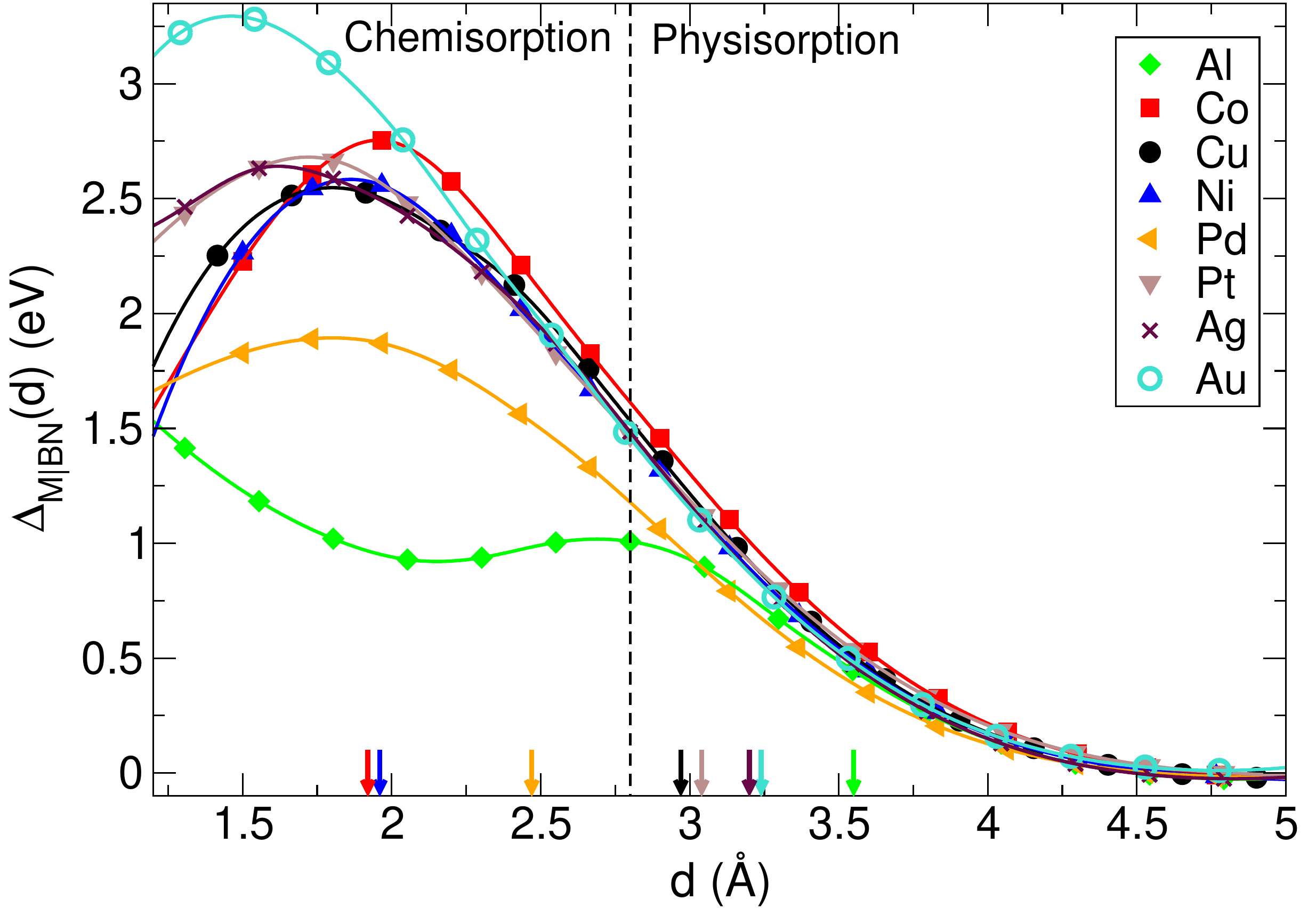}
\caption{(Color online) Potential steps $\Delta_{\rm M|BN}$ at metal$|$\BN{} interfaces as a function of the distance $d$ between the metal surfaces and the $h$-BN plane. The arrows at the bottom indicate the LDA equilibrium binding distances $d_\mathrm{eq}$ of $h$-BN on these metal surfaces, see Table \ref{tbeq}.} 
\label{figE}
\end{figure}

Figure \ref{figE} shows the potential steps $\Delta_{\rm M|BN}$ calculated with $h$-BN in a fixed, planar geometry as a function of separation $d$ for the metals that require a small supercell\cite{fn2}. A distinction is made between chemisorption and physisorption regimes as discussed in the previous paragraph. In the chemisorption regime the curves for different metals can be quite different, which is consistent with the notion that details of the chemical bonding of $h$-BN to a metal substrate should depend on the metal. The potential steps in the chemisorption regime are larger than those given in Table~\ref{tbeq} because we used a fixed $h$-BN structure in calculating Fig.~\ref{figE}. In the chemisorption regime the structure of the $h$-BN layer is perturbed by chemical bonding to the substrate and becomes buckled, which reduces the potential step.

Remarkably, in the physisorption regime the $\Delta_{\rm M|BN}(d)$ curves for the different metals converge; in this regime it is possible to describe all curves with a single function. As the dominant contribution to the potential step in this regime is exchange repulsion between the metal surface and the $h$-BN sheet, and exchange repulsion varies roughly exponentially with distance, a reasonable ansatz for a functional form is an exponential function times a polynomial \cite{Bokdam13}. We find that sufficient accuracy can be obtained using a second order polynomial,
\begin{equation}
\Delta_{\rm M|BN}(d)=e^{-\gamma d}\left(a_0+a_1d+a_2d^2\right).
\label{eq:MBN}
\end{equation}
A least squares fit of this functional form to the $\Delta_{\rm M|BN}(d)$ curves 
for the metals on which $h$-BN is physisorbed (Al, Cu, Ag, Au, Pt) then gives 
$a_0=-1865$ eV, $a_1=1294$ eV/\AA, $a_2=-190.3$ eV/\AA$^2$ and $\gamma=1.85$ 
\AA$^{-1}$. \footnote{The 95\% confidence intervals of the 
parameters $a_{0,1,2}$ are $\pm 291, \pm 184, \pm 29$, respectively. The decay 
parameter $\gamma$ varies between 1.82 \AA$^{-1}$ for Ag and 1.92 \AA$^{-1}$ for 
Pt. The value 1.85 \AA$^{-1}$  gives a reasonable fit for all metals.} 

\begin{figure}[!t]
\includegraphics[scale=0.33]{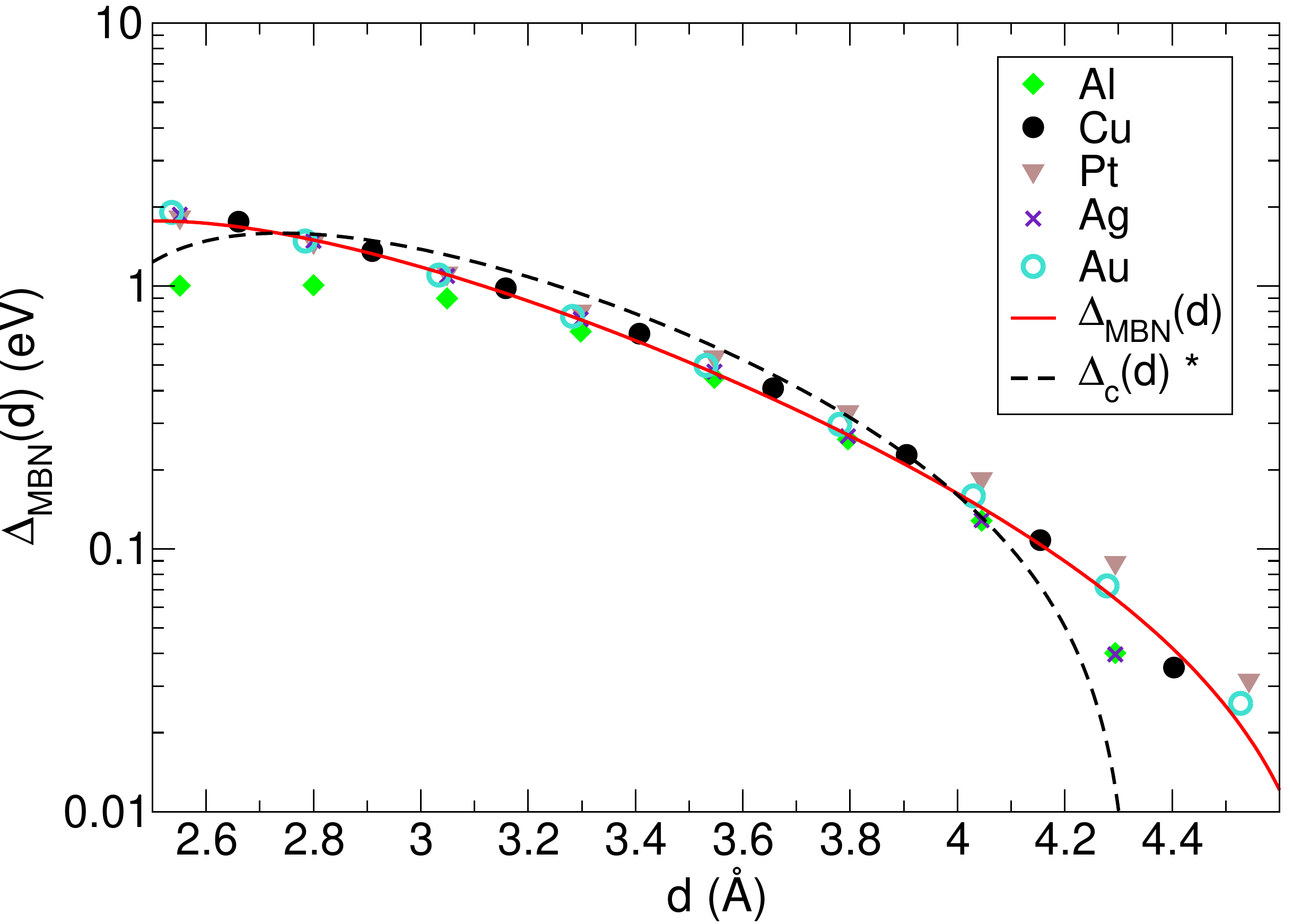}
\caption{(Color online) Potential steps $\Delta_{\rm M|BN}(d)$ at metal$|h$-BN interfaces as in Fig.~\ref{figE} plotted on a logarithmic scale for the metals on which $h$-BN is physisorbed. The solid line gives the fit function of Eq.~\ref{eq:MBN}; the dashed line gives the function $\Delta_c(d)$ used for graphene on these metals in Ref.~\onlinecite{Khomyakov:prb09}.}
\label{figEE}
\end{figure}

This fit function is shown in Fig.~\ref{figEE} along with the data points, which 
shows that the fit is remarkably good. An exception seems to be Al for $d<3.0$ 
\AA, but this is to be expected as there regime chemisorption sets in. The 
deviation of the function from a simple exponential is most prominent for 
$d>4.0$ \AA. The reason for this is that the dipole resulting from exchange 
repulsion decreases rapidly to zero at large distances. What remains is a small 
dipole that results from van der Waals interactions between the metal and $h$-BN 
\cite{Bokdam13} that does not go exponentially to zero as a function of $d$, but 
rather as a power law $d^\alpha$. In the present case, the numbers become too 
small to pinpoint the exact value of $\alpha$.

A function of the form given in Eq.~(\ref{eq:MBN}) was used by Khomyakov \emph{et al.} to describe the potential step encountered in the physisorption of graphene on metal substrates \cite{Giovannetti:prl08,Khomyakov:prb09}. The bonding of $h$-BN to the metals studied in the present paper is very similar. A comparison of the equilibrium binding separations and energies in Table~\ref{tbeq} to those for graphene on metals in Ref.~\onlinecite{Khomyakov:prb09} gives the same classification into chemisorption and physisorption for both graphene and $h$-BN. The $h$-BN layer forms a slightly stronger bond  with the metals than graphene does. In the case of graphene the analysis of the potential step is complicated by the electron transfer between the metal and graphene \cite{Giovannetti:prl08,Khomyakov:prb09,Khomyakov:prb10}. In the case of $h$-BN, this cannot (easily) happen because it is a large band gap insulator. After modeling the effect of electron transfer in the metal$|$graphene case, a contribution to the potential step $\Delta_c(d)$ remained that could be fitted with the same functional form as Eq.~(\ref{eq:MBN}). The function $\Delta_c(d)$ extracted in Refs.~\onlinecite{Giovannetti:prl08,Khomyakov:prb09} is also shown in Fig.~\ref{figEE}. For distances in the typical physisorption regime $2.8 <d <4.0$ \AA, this function is very close to that obtained for metal$|h$-BN interfaces, indicating that in their adsorption on metal surfaces, $h$-BN and graphene behave very similarly.

\subsection{Energy level alignment}\label{ssec:levels}

\begin{figure}[!t]
\includegraphics[width=8.5cm]{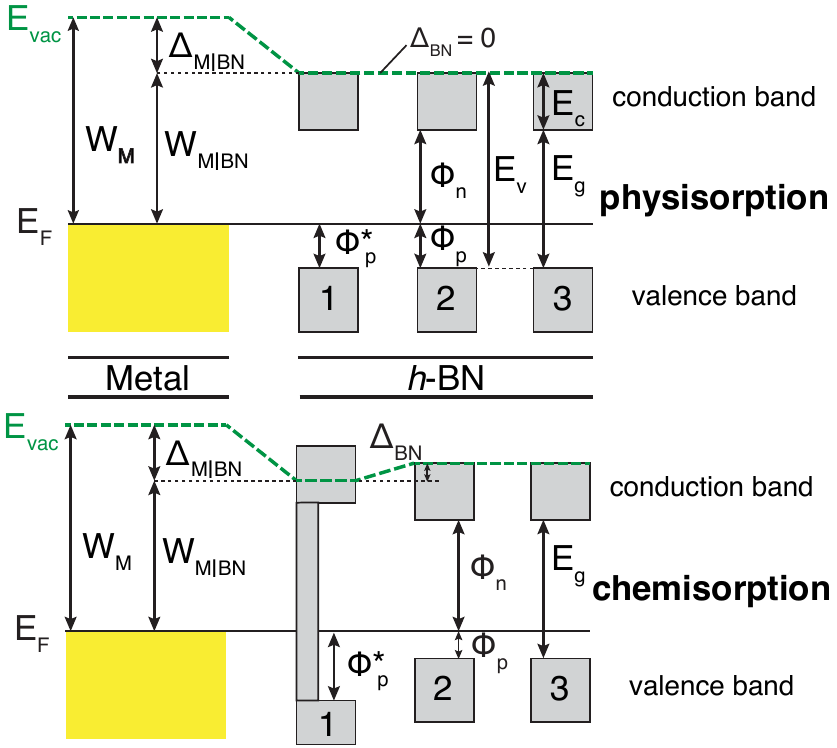}
\caption{(Color online) 
Top: schematic illustration of the energy levels at a metal$|h$-BN interface for three layers of \BN{} when the first layer is physisorbed. $E_\mathrm{vac}$ and $E_F$ are the vacuum level and the Fermi level, respectively. $W_{\rm M}$, $W_{\rm M|BN}$, and $\Delta_{\rm M|BN}$ are the work function of the clean metal surface, that of the metal$|h$-BN structure, and the interface potential step, respectively. $E_v$, $E_c$, and $E_g$ are the top of the valence band of $h$-BN, the bottom of the conduction band, and the band gap, respectively. $\Phi_p$ and $\Phi_n$ are the Schottky barrier heights for holes and electrons, respectively. In the physisorbed case $\Phi^*_p=|E_v-E_F|\approx \Phi_p$. 
Bottom: if the first layer of \BN{} is chemisorbed on the metal (resulting in a distorted $h$-BN layer) a small additional potential step $\Delta_{\rm BN}$ of order 0.1 eV develops between the first and the second \BN{} layer.  In general, $\Phi^*_p$ and $\Phi_p$ can differ by much more, by electron volts (see Tables~\ref{tbeq} and \ref{tbeqvdW}).   
}
\label{figA}
\end{figure}

A schematic illustration of the Schottky barrier formed at a metal$|h$-BN interface is shown in Fig.~\ref{figA} (top). We define the $p$-type Schottky barrier height as 
\begin{equation}
\Phi_p   = E_{v}-W_{\rm M}+\Delta_{\rm M|BN}(d)-\Delta_{\rm BN}.
\label{eq:schottky}
\end{equation} 
Here $W_{\rm M}$ is the work function of the clean metal surface, $\Delta_{\rm 
M|BN}(d)$ is the interface potential step discussed in the previous section, 
$\Delta_{\rm BN}$ is a small potential step between the first and second $h$-BN 
layer (discussed below) and $E_{v}$ is the position of the top of the valence 
band of isolated $h$-BN with respect to the vacuum energy (its ionization 
potential). We define all these quantities as positive numbers, implying that 
the \textit{p}-type Schottky barrier height is a positive number. With $E_g$ the 
band gap of $h$-BN, the \textit{n}-type Schottky barrier height is given by
\begin{equation}
\Phi_n   =E_g -\Phi_p, \label{eq:nschot}
\end{equation}
and is also a positive number. If the first \BN{} layer is chemisorbed on the 
metal surface its electronic properties are altered. As a result, a small 
additional potential step $\Delta_{\rm BN}\approx 0.1$ eV is formed when a 
second \BN{} layer is adsorbed on top of the chemisorbed one, see 
Fig.~\ref{figA} (bottom). A potential step of similar size is found at the 
interface between graphene and \BN{}.\cite{Bokdam:nanol11,Bokdam:prb13} For 
physisorbed \BN{}, the potential step between the first and second \BN{} layers 
is small, i.e., $\Delta_{\rm BN}<0.1$ eV, and can be neglected. Even for 
chemisorbed \BN{}, where $\Delta_{\rm BN}\approx 0.1$ eV, it presents only a 
small correction to $\Phi_{p}$. In all cases there is no potential step between 
the second and third \BN{} layers, so in that sense the second layer already 
resembles a layer in bulk \BN{}. The Schottky barrier heights calculated with 
Eqs.~(\ref{eq:schottky}) and (\ref{eq:nschot}) are given in the last two columns 
of Tables \ref{tbeq} and \ref{tbeqvdW}. 

\begin{figure}
\includegraphics[width=8.5cm]{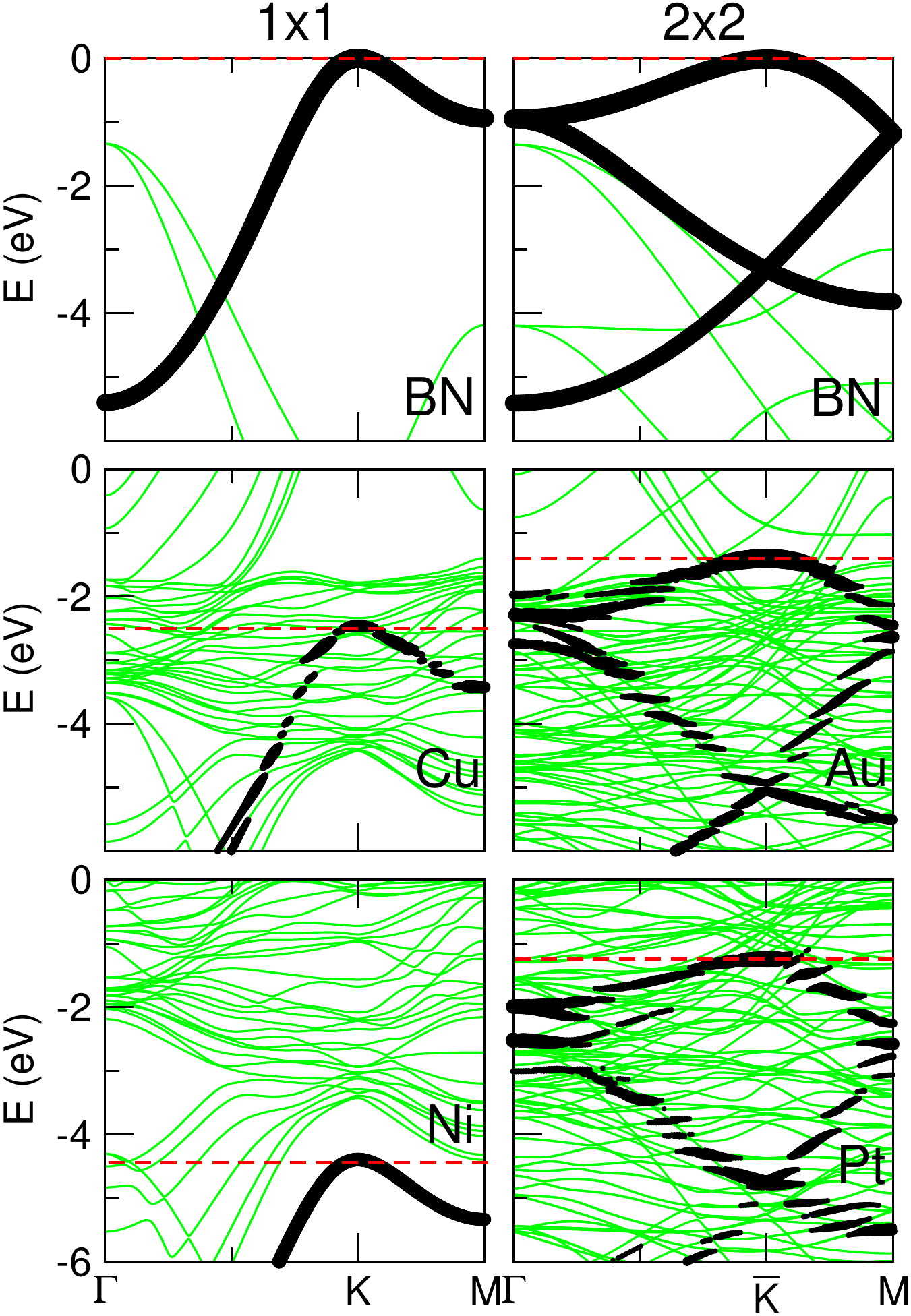}
\caption{(Color online) LDA band structures of a $h$-BN monolayer absorbed on 
the (111) surfaces of Cu, Ni,  Au and Pt. The top two panels show the bands of 
isolated $h$-BN as calculated in a 1$\times$1 unit cell (left) and a 2$\times$2 
supercell (right). The amount of $h$-BN $p_z$ character is indicated by the 
blackness of the bands. The horizontal red dashed lines indicate $\Phi^*_{p}$, 
i.e. the position of the top of the $h$-BN valence band. The zero of energy 
corresponds to the Fermi level. \cite{fn4}
} 
\label{figG}
\end{figure}

Before discussing these results, we examine in more detail how and to what extent the electronic structure of $h$-BN is modified upon forming an interface with a metal. Band structures of metal$|h$-BN interfaces are shown in Fig.~\ref{figG} for some representative metals. If the $h$-BN layer interacts only weakly with the metal substrate, as one expects to be the case for physisorption, then the bands of $h$-BN should still be identifiable. Fig.~\ref{figG} shows that this is indeed the case for a $h$-BN monolayer on Au and Pt(111), for instance. From the band structure one would determine a Schottky barrier height as $\Phi^*_p = |E_v - E_F|$. For a single layer of \BN{} physisorbed on Pt, Ag, Au and Al, $\Phi^*_p$ is within 0.1 eV of the Schottky barrier $\Phi_p$ obtained from Eq.~(\ref{eq:schottky}). 

\begin{figure}
\includegraphics[scale=0.33]{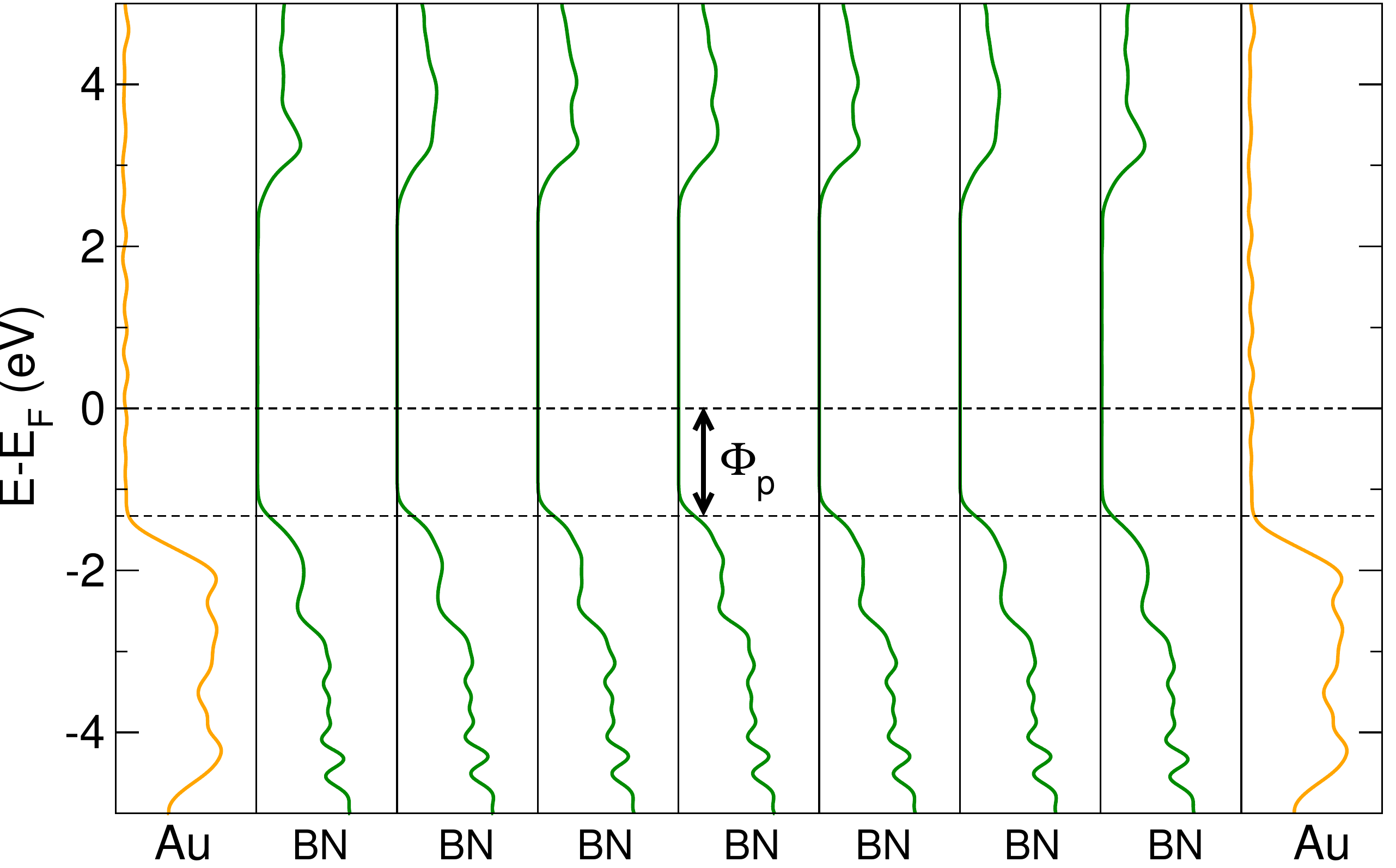}
\includegraphics[scale=0.33]{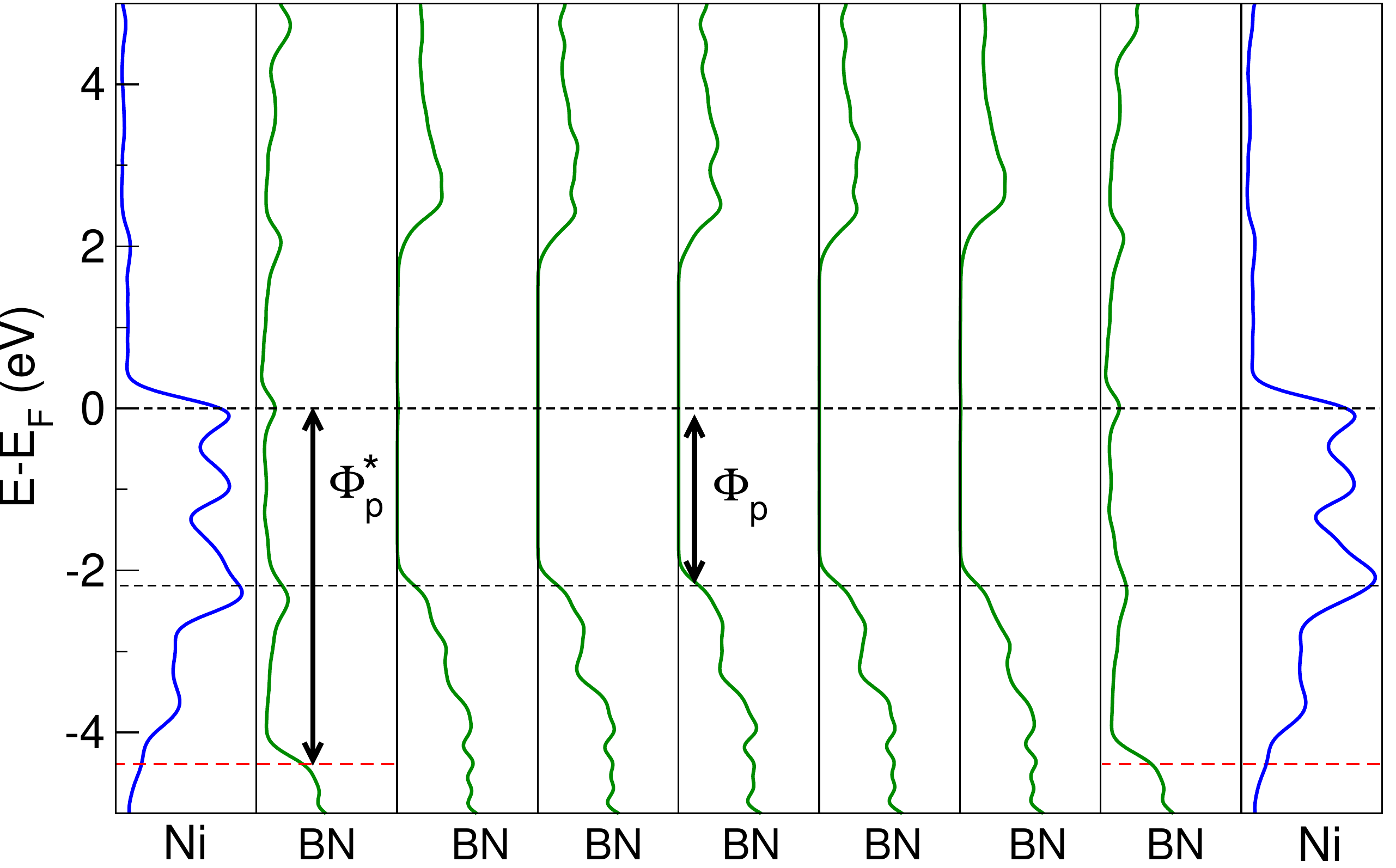}
\caption{(Color online) LDA layer projected densities of states (PDoS)\cite{fn1} for seven layers of $h$-BN sandwiched between Au(111) (top) and Ni(111) (bottom) electrodes. The $p$-type Schottky barrier heights are indicated by $\Phi_p$. } 
\label{figF}
\end{figure}

If the $h$-BN interacts strongly with the metal substrate, i.e., when it is 
chemisorbed, one does not expect such a simple analysis to hold. Fig. \ref{figG} 
also shows the band structure of Ni(111)$|h$-BN, where $h$-BN is strongly bonded 
to the metal surface with a short binding separation, see Table \ref{tbeq}. 
Somewhat surprisingly, the uppermost valence band of $h$-BN can still be clearly 
identified. Contrary to what has been suggested from experiment 
\cite{Nagashima:prl95}, this does {\em not} imply that $h$-BN is physisorbed.  
The identification of a single band of an adsorbed layer is not sufficient to 
decide whether that layer is physisorbed or chemisorbed. In this particular 
case, on the basis of its large binding energy and small equilibrium separation, 
\BN{} is clearly chemisorbed on Ni(111), see Tables \ref{tbeq} and 
\ref{tbeqvdW}.  As shown in Fig.~\ref{figF}, the lower conduction 
bands of the $h$-BN layer in contact with Ni are strongly perturbed by the 
interaction with the substrate. Moreover, from Fig.~\ref{figG}, $\Phi^*_p  > 4$ 
eV, whereas $\Phi_p  = 2.2$ eV, so extracting the Schottky barrier height from 
the band structure of an adsorbed \BN{} monolayer would give a 
misleading result.  

The situation for \BN{} on Cu(111) is less clear. On the one hand, one would classify this as physisorption on the basis of the small binding energy and relatively large equilibrium separation. This is consistent with the observation that the \BN{} bands can be clearly identified for monolayer \BN{} on Cu(111), see Fig.~\ref{figG}. On the other hand, the sizable difference between $\Phi^*_p=2.47$ eV and $\Phi_p=2.10$ eV indicates that the \BN{} bands are perturbed more than is usual in physisorption.  

If the electronic structure of a single $h$-BN sheet is strongly perturbed by adsorption, we need to look at thicker $h$-BN layers to determine the Schottky barrier height. Fig.~\ref{figF} shows the layer projected densities of states (PDoS) for two M$|$($h$-BN)$_n|$M structures with seven ($n=7$) layers of $h$-BN sandwiched between metal slabs. When $h$-BN is physisorbed on the metal, as is the case for a Au(111) substrate (top), the PDoS of all the $h$-BN layers are very similar, even the PDoS of the layers that are in direct contact with the metal substrate. This is consistent with the band structure of $h$-BN being unperturbed by physisorption as discussed above. The Schottky barrier height can be determined from the position of the top of the $h$-BN valence band that is easily identified, being the same for all $h$-BN layers. In this case, we can say that the barrier is localized at the Au$|h$-BN interface. These conclusions hold for all of the metals on which $h$-BN is physisorbed (with, as discussed above, the exception of Cu), i.e., for  Al, Ag, Au, and Pt.

When $h$-BN is chemisorbed (Co, Ti, Ni, Pd), the first layer is perturbed electronically, as well as structurally (buckled), and its interaction with the unperturbed second layer gives rise to an additional small potential step. Only the two layers directly adjacent to the metal are affected and the top of the valence band then remains constant when more $h$-BN layers are added. This means that in the chemisorbed case, the Schottky barrier extends over the $h$-BN monolayer that is adsorbed on the metal. The difference with the physisorbed case is illustrated in the lower panel of Fig.~\ref{figF}, which shows the PDoS for a Ni(111)$|h$-BN$|$Ni(111) structure. The PDoS of the first layer shows signs of (relatively) strong hybridization between the $h$-BN states and the substrate Ni(111) states. However, already the second layer shows a PDoS that is typical of an unperturbed $h$-BN layer. This PDoS allows us to estimate the Schottky barrier height by looking at the top of the $h$-BN valence band. The value obtained is $\Phi_p = 2.19$ eV in good agreement with the 2.20 eV given in Table \ref{tbeq}. 

\subsection{Incommensurable metal$|h$-BN systems}\label{ssec:incom}

In section \ref{ssec:mbinding} we argued that $h$-BN on Cu, Pt and Pd(111) might be expected to be incommensurable, forming superstructures with $m\times m$ $h$-BN unit cells on $n\times n$ metal surface unit cells, with the integers $n$ and $m$ depending on the lattice mismatch and the angle between the $h$-BN and metal lattices. Recent STM experiments show that $h$-BN grown on Cu(111) exhibits moir\'e patterns with periods as large as 14 nm \cite{Joshi:nanol12}. These experiments also show a spread of 0.3 eV in the local work function  depending on the position on the surface, i.e., whether the STM tip is on a ``top'' or a ``valley'' position in the moir\'e pattern.

Incommensurable structures with large supercells are not directly accessible to first-principles calculations. We therefore adopt a simpler approach to estimate the spread in the local work function that might occur in an incommensurable structure. We investigate a fixed orientation of a $h$-BN layer with respect to a Cu(111) substrate, and do not consider general angles. We do however vary the position of the $h$-BN lattice with respect to the substrate. The underlying assumption is that in an incommensurable structure with a sufficiently large period, the bonding locally resembles that of a commensurable structure with a fixed relative displacement of the two lattices. 

\begin{figure}[b]
\includegraphics[width=8.5cm]{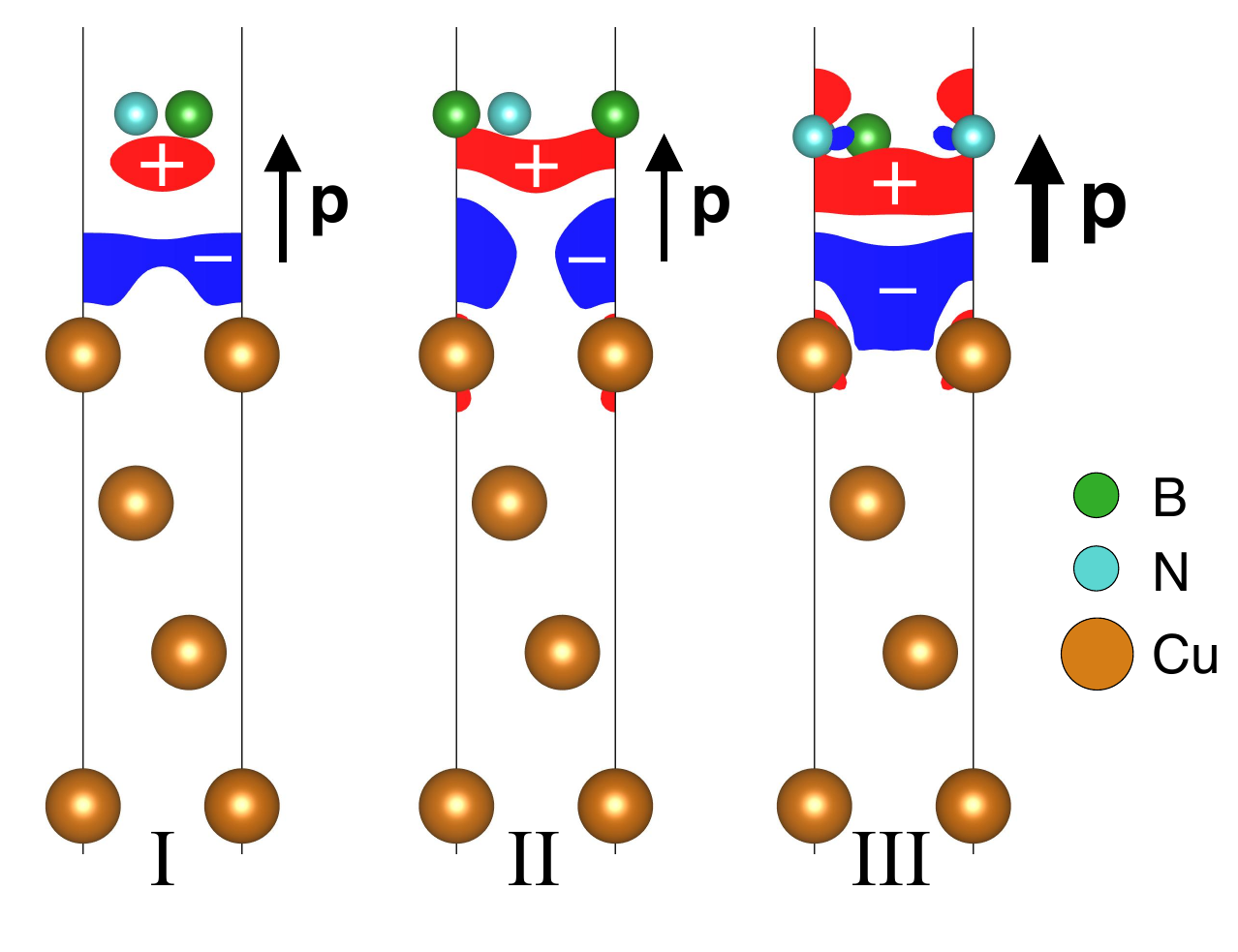}
\caption{(Color online) 
Interface dipole ${\bf p}$ for the three symmetric bonding configurations of commensurable $h$-BN on Cu(111) sketched in Fig.~\ref{figB}(c).
The formation of the interface dipole is illustrated by the charge displacement $-e \Delta n({\bf r})$ (Eq.~\eqref{deltan}) in a plane containing B, C, and N atoms. Blue and red indicate regions of negative and positive charge density, respectively.} 
\label{figH}
\end{figure}

The local bonding can correspond to that of the configurations I, II and III of Fig. \ref{figB}(c), for example. The interface dipole is sensitive to the local bonding. This is clearly visible in Figure \ref{figH}, which shows the charge density difference $-e \Delta n_{\rm Cu|BN}$ for these three configurations. The size of the interface dipole, and therefore of the interface potential step $\Delta_{\rm Cu|BN}$, depends on the configuration, see Table \ref{tbCuBN}. The dipole is largest if the N atoms of $h$-BN are adsorbed on top of the Cu atoms, as in configuration III, and smallest if both B and N atoms are adsorbed at hollow or bridge positions, as in configuration I. The difference in $\Delta_{\rm Cu|BN}$ between these two configurations is 0.45 eV. 

\begin{figure}
\includegraphics[scale=0.45]{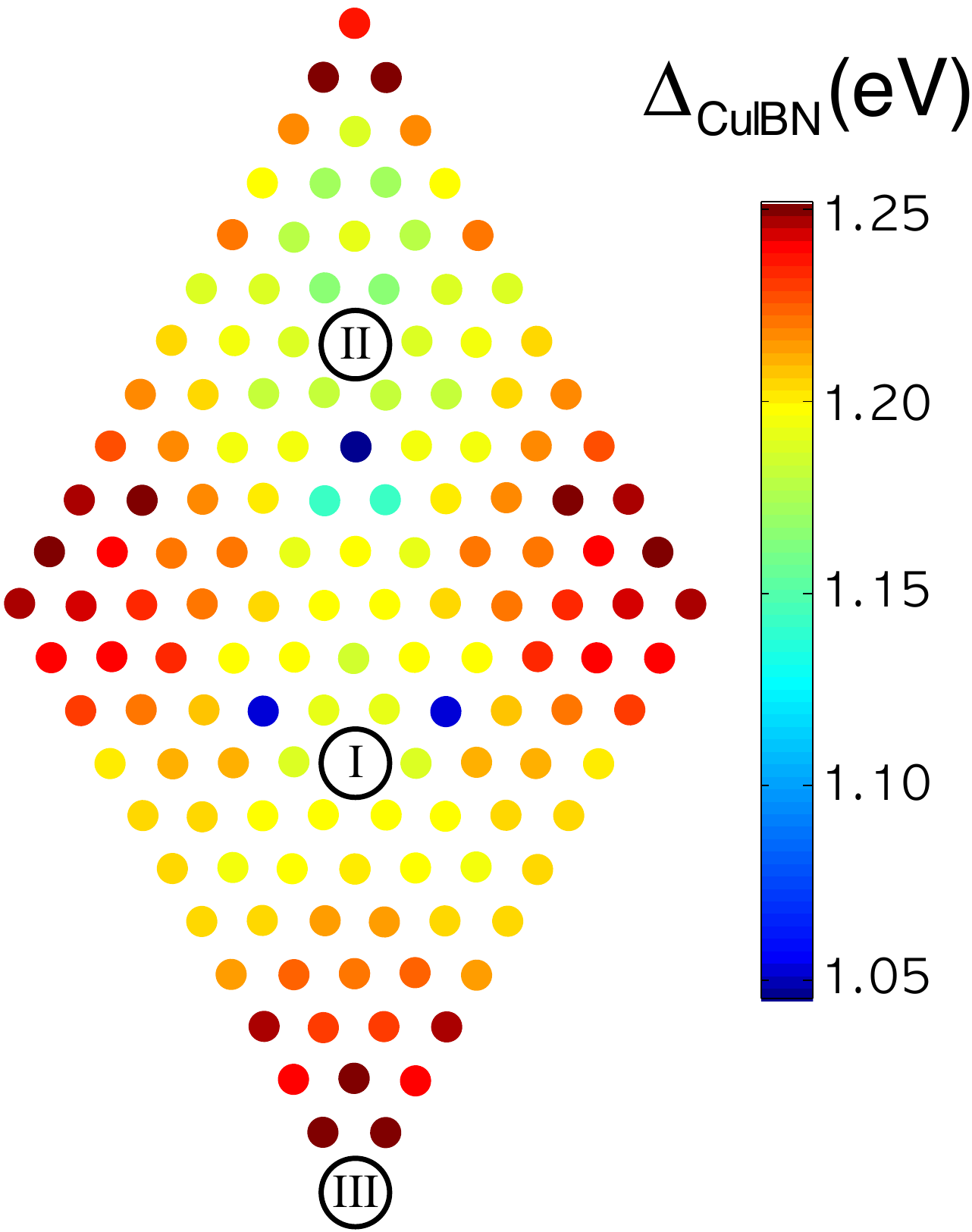}
\caption{(Color online) The potential step $\Delta_{\rm Cu|BN}$ (LDA) at the Cu(111)$|h$-BN interface as a function of the in-plane position of $h$-BN with the Cu(111)--$h$-BN separation fixed at 3.0 \AA. Each dot represents a distinct configuration in steps of $\nicefrac{1}{12}$th of the surface lattice vectors of commensurable $h$-BN on Cu(111).} 
\label{figI}
\end{figure}

The equilibrium separation of $h$-BN from Cu(111) in configuration I is $0.3$ \AA{} larger than in configuration III. Choosing a fixed separation of 3.0 \AA{}, we find a difference in $\Delta_{\rm Cu|BN}$ of 0.1 eV between the two configurations. Scanning the $h$-BN layer over the Cu(111) surface at this fixed separation yields the potential landscape for $\Delta_{\rm Cu|BN}$ shown in Fig.~\ref{figI}. The difference between the extrema of this landscape is 0.2 eV. The interface potential step in an incommensurable structure with a large period varies with the local interface structure, giving rise to a moir\'e pattern in $\Delta_{\rm Cu|BN}$. 

The experimentally observed spread in the local work function is 0.3 eV \cite{Joshi:nanol12}, smaller than the 0.45 eV difference between configurations I or III obtained from fully relaxed homogeneous structures and larger than the 0.2 eV obtained with a constant \BN{}$-$substrate separation. In an incommensurable structure, strain in the $h$-BN layer will prevent the structure from fully relaxing locally to configurations I or III and we expect the experimental result to be a compromise between the two extreme cases we have presented.

\section{Summary and conclusions}\label{sec:discussion}
We calculate the Schottky barrier heights between $h$-BN and a range of metals from first principles, using DFT with vdW-DF and LDA functionals. The close-packed metal surfaces of Al, Ti, Ag, Cu, Ni, Co, Pd, Au, and Pt have work functions ranging from 4.2 to 6.0 eV. When brought into contact with $h$-BN, Schottky barriers for holes are formed with heights $\Phi_p$ ranging from 1.2 to 2.3 eV (vdW-DF: 0.9 to 2.4 eV). There is, however, no simple relation between the size of the work function and that of the Schottky barrier height. A potential step at the interface, $\Delta_{\rm M|BN}$, caused by the interaction between $h$-BN and the metal makes a major contribution to the Schottky barrier height, cf. Eq. \ref{eq:schottky}. 

The interaction between $h$-BN and different metals falls into two distinct categories. $h$-BN is chemisorbed on Co(0001), Ti(0001), Ni(111) and Pd(111) surfaces with equilibrium separations $\lesssim 2.5$ \AA. On the (111) surfaces of Al, Cu, Ag, Au, and Pt, $h$-BN is physisorbed with equilibrium separations $\gtrsim 3.0$ \AA. Chemisorption results in large interface potential steps that depend on the details of the chemical interaction; for the examples we considered, $\Delta_{\rm M|BN}$ is 0.8-1.8 eV (vdW-DF: 0.9-1.9 eV). Physisorption gives somewhat smaller interface potential steps; we found values of $\Delta_{\rm M|BN}$ ranging from 0.4-1.2 eV (vdW-DF: 0.4-0.8 eV). In the physisorption regime, the dependence of the potential step on the distance $d$ from the $h$-BN plane to the metal surface can be described reasonably well by a single function, Eq. \ref{eq:MBN}, which is independent of the metal. 

The range of values of the potential step observed for the different metals then reflects the range of the equilibrium binding distances of $h$-BN on these different metals. The differences between the potential steps calculated with vdW-DF and LDA can also be ascribed to differences in equilibrium binding distances obtained with these two functionals. The interface charge distributions obtained with vdW-DF and LDA are very similar, both in the physisorption, as well as in the chemisorption regimes. 

In our calculations, of necessity we force the metal$|h$-BN interface into a commensurable structure. Such a structure is likely if the mismatch between the $h$-BN and the metal surface lattices is negligible, or if the metal--$h$-BN interaction is strong. However, the lattice mismatch between $h$-BN and Cu, Pd, Pt(111) is $\gtrsim 3$\%, and the metal--$h$-BN binding is not sufficiently strong to overcome the strain energy required to compress or stretch $h$-BN to match the metal lattice. Therefore in these cases the metal$|h$-BN interface forms an incommensurable structure. Assuming that the period of the resulting superstructure is sufficiently large, one can determine an in-plane dependence of the interface potential step from calculations on commensurable systems. For $h$-BN on copper we find an in-plane potential variation in the range 0.2-0.45 eV. This leads to an in-plane variation of the Schottky barrier height of the same magnitude. 

In the chemisorbed cases, the structures are likely to be commensurable. The adsorbed $h$-BN monolayer is then perturbed electronically as well as structurally, i.e., the layer becomes buckled with a buckling amplitude $\sim 0.1$ \AA. The interaction between this perturbed layer and a second $h$-BN layer leads to an additional potential step, whose size however is limited to $\sim 0.1$ eV.

\begin{table*}[t]
\caption{Calculated (LDA) and experimental potential steps 
$\Delta_{\rm M|BN}$ 
at the metal$|h$-BN interface; (average) equilibrium separation $d_\mathrm{eq}$ 
between the metal surface and the $h$-BN plane, using the structures shown in 
Fig.~\ref{figB}, starting from in-plane Pd and Pt lattice constants of 4.76 and 
4.81 \AA, respectively, and compressing the $h$-BN lattice accordingly, see also 
Table \ref{tbeq}; binding energies $E_{\rm b}$ per BN; calculated and 
experimental work functions $W_{\rm M|BN}$ of metal covered with a single sheet 
of $h$-BN and $W_{\rm M}$ of the clean metal surface. The last two columns show 
the Schottky barrier height $\Phi_p$ for holes calculated for several layers of 
$h$-BN, see Eq. (\ref{eq:schottky}) with $E_v=6.09$ eV and the Schottky barrier 
height $\Phi_n$ for electrons, see Eq. (\ref{eq:nschot}), where we have used the 
experimental band gap $E_g = 5.97$ eV.\cite{Watanabe:natm04} 
}
\begin{ruledtabular}
\begin{tabular}{lccccccc|cc}
 M  & $\Delta_{\rm M|BN}$
            & $d_{\rm eq} $
                     & $E_{\rm b}$ 
                                &  $W_{\rm M|BN} $ 
                                       &$W_{\rm M|BN}^{\rm exp} $ 
                                              &  $W_{\rm M} $ 
                                                                  &$W_{\rm M}^{\rm exp} $ 
                                                                                                 & $\Phi_{p}$ 
                                                                                                          & $\Phi_{n}$\\
 \hline
 
 Pd & 1.29 & 2.79    & $-0.106$ & 4.37 & 4.0$^{\rm a}$ & 5.66 & 5.6 $^{\rm b}$ & $1.62$ & 4.35 \\
 Pt & 0.95 & 3.17    & $-0.082$ & 5.15 & 4.9$^{\rm a}$ & 6.10 & 6.1 $^{\rm f}$ & $0.94$ & 5.03 \\
    & (eV) & (\AA{}) &  (eV)    & (eV) &  (eV)         & (eV) &  (eV)          & (eV) & (eV)         \\
\end{tabular}
\end{ruledtabular} 
\newline
$\rm ^a$ Ref. \onlinecite{Nagashima:prl95},
$\rm ^b$ Ref. \onlinecite{Michaelson:jap77}, 
$\rm ^f$ Ref. \onlinecite{Derry:prb89}
\label{tappendix}
\end{table*}

Though the calculated work function of Pt and the $h$-BN ionization potential are very close ($\sim 6$ eV), the calculated Schottky barrier is $\Phi_p \approx 1$ eV. This is caused by a potential step $\Delta_{\rm M|BN}$ of that size, which is formed upon adsorption of $h$-BN on Pt, which effectively lowers the Pt work function. Such potential steps are formed when $h$-BN is adsorbed on all of the metal substrates studied in this paper, which implies that all Schottky barrier heights $\Phi_p \gtrsim 1$ eV. The Schottky barrier heights for electrons are even higher, i.e., $\Phi_n \gtrsim 3.5$ eV. This means that it will not be easy to apply $h$-BN as a semiconductor material. It also means, however, that $h$-BN is a good insulator for use in graphene electronics.

\acknowledgements{M.B. acknowledges support from the European project MINOTOR, 
grant no. FP7-NMP-228424. M.I.K. acknowledges finacial support from the 
European Union Seventh Framework Programme under grant agreement No. 604391
Graphene Flagship. The use of supercomputer facilities was sponsored by 
the Physical Sciences division of the Netherlands Organization for Scientific 
Research (NWO-EW).}

\appendix*
\section{Lattice Mismatch}
We argued in Sec.~\ref{ssec:mbinding} that a metal-BN interface is very likely 
to be incommensurable if $h$-BN is physisorbed and the metal-BN lattice mismatch 
is substantial. When forced to use commensurable structures in electronic 
structure calculations to model incommensurable systems, care must be taken to 
ensure that this does not alter the electronic structure unacceptably. Modifying 
the in-plane lattice constant of a close-packed metal surface by a few percent 
only affects its electronic properties mildly. For instance, the work function 
typically changes on the scale of 0.1 eV, the binding energy of a physisorbed 
species on the scale of 0.01 eV/atom. By contrast, changing the lattice 
parameter of graphene by a few percent can change the way it binds to a metal 
substrate qualitatively. For example, stretching graphene by 4\% to match the 
Cu(111) lattice leads to an unrealistically strong chemisorption of graphene, 
decreasing the binding distance to the Cu surface from 3.3 {\AA} 
\cite{Khomyakov:prb09} to 2.2 \AA. 

For these reasons we chose in Sec.~\ref{ssec:mbinding} to fix the lattice 
parameter of $h$-BN at its equilibrium value and adapt the metal in-plane 
lattice constants accordingly. The adjustments are largest for  Pt(111) and 
Pd(111), whose lattices have to be stretched by 3.5\% and 4.6\%, respectively, 
if a $\sqrt{3} \times \sqrt{3}$ metal surface cell is matched to a $2 \times 
2$ $h$-BN cell. Table \ref{tappendix} gives the results for the 
case where we fixed the in-plane lattice constants of Pt(111) and 
Pd(111) to the experimental values, and compressed the $h$-BN lattice 
accordingly. A comparison with the numbers given in Table \ref{tbeq} shows that 
for this case the differences are moderate. The LDA work functions of the clean 
Pt and Pd(111) surfaces differed there from experiment by ~0.12 eV. Adapting the 
$h$-BN lattice to the experimental metal lattice constants increases the 
adsorption energies by 0.057 eV/BN for Pd and by 0.018 eV/BN for Pt. The 
interface potential step decreases by 0.09 eV for Pt, and increases by 0.04 eV 
for Pd. These differences are similar or smaller than the uncertainties caused 
by using different functionals.


\begin{thebibliography}{60}
\bibitem [{fn2()}]{fn2}%
  \BibitemOpen
  \href@noop {} {}\bibinfo {note} {Because of the large computational cost of
  the Ti$|h$-BN supercell, only LDA calculations were carried out and the
  separation dependence of the interface dipole was not studied.}\BibitemShut
  {Stop}%
\bibitem [{fn1()}]{fn1}%
  \BibitemOpen
  \href@noop {} {}\bibinfo {note} {The PDoS in Figure \ref{figF} is broadened
  by smearing (2nd order Methfessel-Paxton, $\sigma=0.2$ eV). The Schottky
  barrier height is determined in the PDoS calculated with the tetrahedron
  method, which gives a sharp (non-smeared) valence band maximum.}\BibitemShut
  {Stop}%
\end{thebibliography}
\end{document}